\documentclass[aps,pra,twocolumn,nofootinbib,superscriptaddress,showpacs]{revtex4}
\usepackage{amsmath}
\usepackage{amsfonts}
\usepackage{amssymb}
\usepackage{graphicx}
\usepackage{comment}
\usepackage{color}

\begin{document}

\title{Indistinguishable Encoding for Bidirectional Quantum Key Distribution: Theory to Experiment}

\author{J. S. Shaari}
\affiliation{Department of Physics, International Islamic University Malaysia (IIUM),
Jalan Sultan Ahmad Shah, Bandar Indera Mahkota, 25200 Kuantan, Pahang, Malaysia}
\affiliation{Institute of Mathematical Research (INSPEM), University Putra Malaysia, 43400 UPM Serdang, Selangor, Malaysia.}
\author{Suryadi}
\affiliation{Department of Physics, International Islamic University Malaysia (IIUM),
Jalan Sultan Ahmad Shah, Bandar Indera Mahkota, 25200 Kuantan, Pahang, Malaysia}

\date{\today}

\begin{abstract}
We present for the first time, a bidirectional Quantum Key Distribution protocol with minimal encoding operations derived from the use of two `nonorthogonal' unitary transformations selected from two mutually unbiased unitary bases; which are indistinguishable in principle for a single use. Along with its decoding procedure, it is a stark contrast to its `orthogonal encoding' predecessors. Defining a more relevant notion of security threshold for such protocols, the current protocol outperforms its predecessor in terms of security as the maximal amount of information an eavesdropper can glean is essentially limited by the indistinguishability of the transformations. We further propose adaptations for a practical scenario and report on a proof of concept experimental scheme based on polarised photons from an attenuated pulsed laser for qubits, demonstrating the feasibility of such a protocol.
\end{abstract}

\pacs{03.67.Dd}
\keywords{Quantum cryptography}

\maketitle

\section{Introduction}
Quantum cryptography, or more specifically quantum key distribution (QKD) provides for a solution to the courier problem of distributing secret keys between two parties to be utilised for a one-time pad cryptographic protocol. Arguably a first rather direct practical application of quantum physics, with realisations mainly in terms of optical based implementations, QKD's security is guaranteed by physical laws and saw its debut in the famous BB84 \cite{BB84} protocol, where one party, commonly referred to as Alice would send a photon prepared in one of two bases (mutually unbiased) over a quantum channel to another, Bob, for his measurements to determine the state sent. This is a straightforward scenario of `prepare and measure'; i.e. Alice prepares a quantum state while Bob measures. An adversary, Eve, would not be able to determine the states sent without inducing any errors. In general, channels are noisy and detection alone in any QKD protocol would prove to be impractical. More importantly, an estimation of the amount of information Eve may have gleaned can be inferred from the error between Alice and Bob and thus, below a prescribed threshold, a secret key can nevertheless be distilled by the legitimate parties. This is done by first correcting any errors between them using error corrections (EC) codes and Eve's knowledge of the key can be reduced to arbitrarily low levels using privacy amplification (PA) procedures (we refer to \cite{Gisin, HKL} for excellent reviews on quantum cryptography).  

While variants of the first QKD protocol has seen much development, a departure from a prepare and measure scenario was imagined in a QKD protocol making bidirectional use of the quantum channel between Alice and Bob, sometimes referred to as two-way QKD (we shall use the terms bidirectional and two-way interchangeably). It was first reported in 2003 in \cite{Bostrom} and later saw its evolution into various forms, improving on security and some on practicality \cite{Qin,FG,LM05}. The essential feature of the protocol is the encoding of information by one party, Alice, by executing unitary transformations on qubits received from (and prepared by) another party, Bob who would later measure it in the same basis he prepared it in. Information bits for secret key generation is derived from the different transformations, in clear contrast to BB84 like schemes where information for the same purpose is simply the state themselves. Hence, while the latter requires an eavesdropper to estimate the state of a traveling qubit between the legitimate parties for successful eavesdropping, two-way protocols challenge eavesdroppers to estimate the evolution of an unknown state as it travels to and fro between Alice and Bob. Practical implementations realised include those reported in \cite{Cere} with entangled photons, \cite{Khir1,Khir2} with weak (attenuated) pulsed laser as photon sources and even using telecommunication wavelengths in \cite{Rupesh} to cite a few.

However, in all these, the unitary transformations had mostly been limited to those that may be described by the Pauli matrices, $X,Y,Z$ plus the identity operator $\mathbb{I}$ (in most cases, it would be the $iY$ and the identity $\mathbb{I}$) and the security of the protocols mainly lie in the inability to determine conclusively the traveling qubits' states randomly prepared by Bob in one of two mutually unbiased basis (MUB). Bob himself obviously can distinguish between the transformations; 
in principle, these unitary transformations can be distinguished perfectly even for a single use \cite{dariano}. Ref.\cite{bisio} referred to these transformations as a set of \textit{orthogonal unitaries}. While this on its own does not hinder an eavesdropper (Eve) to ascertain without ambiguity the transformations executed by Alice, it does result in the former introducing errors should measurements be made instead on the received states. To this effect the legitimate party would, randomly interlude their encoding/ decoding runs with prepare and measure runs where Bob's prepared states are, with a certain probability, measured by Alice, akin to a BB84 scenario and is referred to as the \textit{control mode} (CM). The encoding/ decoding runs are denoted as the \textit{encoding mode} (EM).  

While being operational and secure, these protocols in some sense betray the essence of utilising physical laws directly affecting Eve's ability to eavesdropping the encodings. As prepare and measure schemes delivers by capitalising on imperfect state estimations, these two-way schemes unfortunately do not, in an analogous way, prescribe imperfect estimation of unitary transformation as part of its working engine. The idea of Alice actually using transformations which are in principle indistinguishable for a single use for encoding purpose was noted in \cite{dar2} where the unitaries would be selected randomly from two mutually unbiased bases of orthogonal qubit unitaries (a term used in \cite{bisio}). A study, more focused towards QKD for such two-way protocols using qubits selected from 2 MUBs encoded with such indistinguishable `nonorthogonal unitaries' was reported in \cite{jss}. The term nonorthogonal unitaries can be traced to  Ref. \cite{non} and the relevant definition was later given in \cite{jss}. A proper formalization for the structure of such unitaries was studied as mutually unbiased unitary bases (MUUB) in \cite{jssarxiv}. In short, two orthogonal bases, $\mathcal{B}_0$ and $\mathcal{B}_1$, for some $n$-dimensional subspace of $2\times 2$ matrices are defined as sets of MUUB when 
\begin{eqnarray}
\left|\text{Tr}(B_0^{i\dagger}B_1^j)\right |^2=C~~,~~\forall B_0^i\in\mathcal{B}_0, B_1^j\in\mathcal{B}_1, 
\end{eqnarray}
for $ i,j,=1,\ldots, n$ and some constant $C\neq 0$; $C$ equals $1$ and $2$ for $n=4$ and $n=2$ respectively \cite{jssarxiv}. 

In this work, we describe and analyse a bidirectional QKD protocol which uses a minimal number of indistinguishable unitary for encodings where each encoding is selected from two different MUUB. Given the use of only 2 unitary operators, differently from \cite{jss} (which used 4) as well as other two-way QKDs, the very decoding procedure by Bob would be radically different from previously reported two-way protocols. Beginning with an ideal protocol, we brief on its merits in a depolarising channel and provide a security analysis which demonstrates its clear advantage over its predecessor, the protocol of \cite{LM05}. We further report on an experimental proof of concept for the protocol revealing its feasibility. 

\section{Bidirectional QKD with Two Mutually Unbiased Unitaries}
\noindent 
The protocol is based on the same bidirectional use of the quantum channel where Bob sends to Alice a qubit prepared in a basis of his choice. Alice would then encode using one of two unitary transformation before submitting to Bob for his measurements. We consider the case where Alice uses unitaries described as rotations around the $y$ axis of the Poincare sphere given respectively by 
\begin{eqnarray}
R_y(\zeta)=\cos{\left(\dfrac{\zeta}{2}\right)}\mathbb{I}-i\sin{\left(\dfrac{\zeta}{2}\right)}Y.
\end{eqnarray}
We choose only two angles for $\zeta$ in this work, namely, $\zeta=0$ (corresponding to a passive operation) and $\zeta=-\pi/2$ which corresponds to flipping states between the mutually unbiased orthonormal $X$ and $Z$ bases. The transformations are in fact elements taken from either of two sets of MUUB \cite{jss}, $\{I,Y\}$ and $\{R_y(\pm\pi/2)\}$ with respect to one another, 
 \begin{eqnarray}
\left|\text{Tr}(IR_y(\pm\pi/2))\right |^2=\left|\text{Tr}(Y^\dagger R_y(\pm\pi/2))\right |^2=2.
\end{eqnarray}
 The indistinguishability of these two transformations can be seen from the indistinguishability of an input state for the transformation, from its output. With an arbitrary state $|\psi\rangle=\cos{(\theta/2)}|0\rangle+\exp{(i\phi)}\sin{(\theta/2)}|1\rangle$, we can quickly observe that the overlap
 \begin{eqnarray}
 |\langle \psi|I R_y(-\pi/2)|\psi\rangle|^2=\dfrac{1}{2}+\dfrac{\sin^2{(\theta)}\sin^2{(\phi)}}{2}
 \end{eqnarray}
has a minimum value of $1/2$. This minimum value is in fact the square of the inner product for two states coming from two MUB. 

Any state lying on the equator of the Poincare sphere would hence provide for minimal overlap. Thus we let Bob prepare a state randomly selected from the basis defined by $\{|0^q\rangle=\cos{(\theta/2)}|0\rangle+\sin{(\theta/2)}|1\rangle,~|1^q\rangle=\sin{(\theta/2)}|0\rangle-\cos{(\theta/2)}|1\rangle\}$ to be submitted to Alice for her encoding (transformation). The value for the angle $\theta$ is also a random choice by Bob.

Once Alice has executed her transformation, the resulting state would be forwarded to Bob for which he shall commit to a measurement in either the same basis he prepared or one rotated by $\pi/2$. Writing Alice's transformation as $U_A$ and Bob's prepared and resulting measured state as $|\psi_f\rangle$ and $|\psi_b\rangle$ respectively, Bob can only determine Alice's encoding conclusively if $\langle\psi_b |U_A|\psi_f\rangle=0.$\footnote{admittedly, this is inspired very much by the SARG protocol \cite{SARG}} As an example, if Bob prepares the computational state $|0\rangle$ and his measurement result is $|1\rangle$, then he knows for certain that Alice could not have used the $\mathbb{I}$ operation. Or, if a measurement had been made in the $X$ basis instead and yields $(|0\rangle+|1\rangle)/\sqrt{2}$, then Bob can infer that Alice had not used $R_y(-\pi/2)$. Bob shall then announce publicly all inconclusive results to be discarded. Assigning the logical value `0' to $\mathbb{I}$ and `1' to $R_y(-\pi/2)$, Alice and Bob can share a key only for $1/4$ of the total qubits used.


\subsection{Security Analysis}

Taking the conventional approach to security analysis of bidirectional QKDs, we consider Eve's strategy is to attack the qubits en route twice, once in the forward path (from Bob to Alice) and once in the backward path (from Alice back to Bob). In the individual paths, the density operator for Bob's qubit, $\rho_B$, on its own does not provide for any information; a qubit prepared as either of the orthogonal states in any basis in the forward path is completely mixed as is the case in the backward path after Alice's encoding.

We shall analyse the protocol based on the methods of \cite{hua,ivan}, where we shall consider each of Bob's traveling state to independently undergo identical interaction with Eve's ancilla prior to Alice's encoding. Then we allow Eve to have access to the entire state in the backward path (after encoding) to extract information and we set no constraint on how she may do this. This is ultimately a very pessimistic stand and provides for a collective attack scenario.

We do however, reasonably require Eve's strategy to simulate a depolarising channel like \cite{ivan}, i.e. Bob's qubit, irrespective of the basis chosen should experience the same amount of noise, essentially undergoing a symmetric attack \cite{Gisin}. Also, like \cite{ivan}, we shall begin with Bob's state in one basis only, and then show that the information Eve should gain for any of Bob's choice of basis is the same. We begin by writing the interaction between Eve's ancillae, $|\mathcal{E}\rangle$, and the travelling qubit (in the computational basis for simplicity) in the forward path as; 
\begin{eqnarray}\label{nonortho}
U|b\rangle|\mathcal{E}\rangle=|b\rangle |\mathcal{E}_{bb}\rangle+|b^\bot\rangle |\mathcal{E}_{bb^\bot}\rangle
\end{eqnarray}
with $b\in\{0,1\}$ and $\bar{0}=1$ ($\bar{1}=0$). Unitarity of the interaction necessitates 
\begin{eqnarray}
\langle \mathcal{E}_{b\bar{b}}|\mathcal{E}_{b\bar{b}}\rangle+\langle \mathcal{E}_{bb}|\mathcal{E}_{bb}\rangle=1,~\langle \mathcal{E}_{bb}|\mathcal{E}_{\bar{b} b}\rangle+\langle \mathcal{E}_{b\bar{b}}|\mathcal{E}_{\bar{b} \bar{b}}\rangle=0 
\end{eqnarray}
and we let $\langle \mathcal{E}_{bb}|\mathcal{E}_{bb}\rangle=F$ and $\langle \mathcal{E}_{b\bar{b}}|\mathcal{E}_{b\bar{b}}\rangle=Q$ and $F+Q=1$. It is also worth noting that, with proper choices for phases, one can ensure all of Eve's scalar products are reals \cite{Gisin}. Now, the value $Q$ is really the probability of Bob's state being measured as one orthogonal to which he sent. Admittedly, in the current protocol, we have not defined the protocol to measure the qubit in the forward path, thus making $Q$ inaccessible; it can easily be determined if we include some form of CM similar to that of \cite{LM05}. We shall return to this point later. 

The state of the system (Bob's qubit after Eve's attack in the forward path) subsequent to Alice's encoding can be written as
\begin{eqnarray}
\rho_{BE}=\dfrac{1}{2}\left[U\rho_BU^\dagger+R_y^E(U\rho_BU^\dagger)R_y^{E\dagger}\right]
\end{eqnarray}
where $\rho_B=\mathbb{I}/2\otimes|\mathcal{E}\rangle\langle \mathcal{E}|$ and $R_y^E=R_y(-\pi/2)\otimes \mathbb{I}_E$ with $\mathbb{I}_E$ being the identity on Eve's Hilbert space. Eve access to the state on the backward path provides her with information of the key, $I_E$, which is given by $S(\rho_{BE})-1$ \cite{ivan} where $S(\rho)$ is the von Neuman entropy given by $-\text{tr}\rho\log_2{\rho}$ for a state $\rho$, which written in terms of its eigenvalues, $\lambda_i$, and eigenkets, then $S(\rho)=-\sum_i \lambda_i\log_2{\lambda_i}$. 

In ascertaining the eigenvalues of $\rho_{BE}$, we adopt the method in \cite{ivan} by calculating the eigenvalues of its Gram matrix representation, $\textbf{G}^{BE}$ \cite{gram}. 
 The eigenvalues, for $\textbf{G}^{BE}$ (which are equal to those of $\rho_{BE}$ including its multiplicities, each being 2) are given by
\begin{eqnarray}
\lambda_\pm=\dfrac{1}{8}(2\pm \sqrt{2(F\cos{x}-Q\cos{y})^2+2}).
\end{eqnarray}
Eve's information gain, $I_E$ can then be written as 
\begin{eqnarray}\label{IE}
I_{E}=S(\rho_{BE})-1=h\left (\dfrac{2-a\sqrt{2}}{4}\right)
\end{eqnarray}
with $a=\sqrt{(F\cos{x}-Q\cos{y})^2+1}$ and $h(x)=-x\log_2{x}-(1-x)\log_2{(1-x)}$ being the Shannon binary entropic function. 

Now, we should stress the fact that Eve's information gain is actually the same for any state Bob could send (constrained to those on the equator of the Poincare sphere) and thus this analysis is valid in considering the protocol as described above where Bob can send any such states. We demonstrate this fact briefly in the section on Methods. Insisting on the same disturbance for any such state sent by Bob, the value for $Q=1-F$ would be given by \cite{Gisin} \footnote{this is relatively easy to derive for any state on the equator of the Poincare sphere and also given in the Methods section.}
and we can eventually write 
\begin{eqnarray}
1-2Q(1+\cos{y})=F\cos{x}-Q\cos{y}.
\end{eqnarray}
Hence, $a=\sqrt{[1-2Q(1+\cos{y})]^2+1}$. We can immediately observe that Eve's best strategy to maximise her information would be to maximise $\cos{y}$; ensuring $Q$ be kept minimal. Thus for a fixed $Q$, let $\cos{y}=1$ (which then fixes $x$) and her information would be given by 
\begin{eqnarray}
I_{E}=h\left [\dfrac{1}{2}\left(1-\dfrac{\sqrt{(1-4Q)^2+1}}{\sqrt{2}}\right)\right].
\end{eqnarray}
It's evident that Eve achieves maximum information, approximately $0.6$ when $Q=0.25$ and is equal to the von Neumann entropy for a mixture of two states derived from two MUBs. 


\subsection{Security Thresholds}

Unlike its prepare and measure cousins, the secret key rate for two way protocols would depend on $2$ parameters of errors, namely the error in the forward path, $Q$, which informs the legitimate parties of Eve's gain, thus the rate for PA, and the error in the backward path, $Q_{AB}$ which tells of the cost in bits for error correction purposes. There is no reason \textit{a priori} to imagine that these two parameters are linked by some straightforward mathematical relationship (even if we assume both channels as depolarising). Hence, the cases of `correlated channels' or `independent channels' as studied in \cite{norm} are really specific models which may (or not) be true in the case of an actual implementation. A `security threshold' can only be determined after both  $Q$ and $Q_{AB}$ are determined. 
More importantly, the notion of security threshold, which is commonly understood as a point denoting the value for error in the channel such that beyond it no secure key can be extracted must give way to the idea of \textit{curves} in a plane defined by $Q$ and $Q_{AB}$ separating regions where key extraction is possible and otherwise.  We define hence, a security threshold as the area for region of the said plane where secret key extraction is possible; i.e. where the secret key rate is greater than zero (we take the maximum values for $Q$ and $Q_{AB}$ for the total region as where Eve's information gain is maximum and Alice-Bob's mutual information is minimal respectively). Secret key rates can be written as $1-I_E-h(Q_{AB})$ \cite{devetak}.

In order to have an idea of the protocol's merit, we compare it to the earlier `orthogonal' protocol of \cite{LM05}, (in some literature referred to as LM05)\footnote{we consider this as fair comparison given the two has essentially identical topologies as well number of states and transformations used}. We calculate and compare the secure key rates per raw key bit for varying values of both $Q$ and $Q_{AB}$. Within the depolarising channel framework, Eve's gain for LM05 is given as $h(1-2Q)$ \cite{hua,ivan}. 
\\
\begin{figure*}
\centering \includegraphics[width=\textwidth]{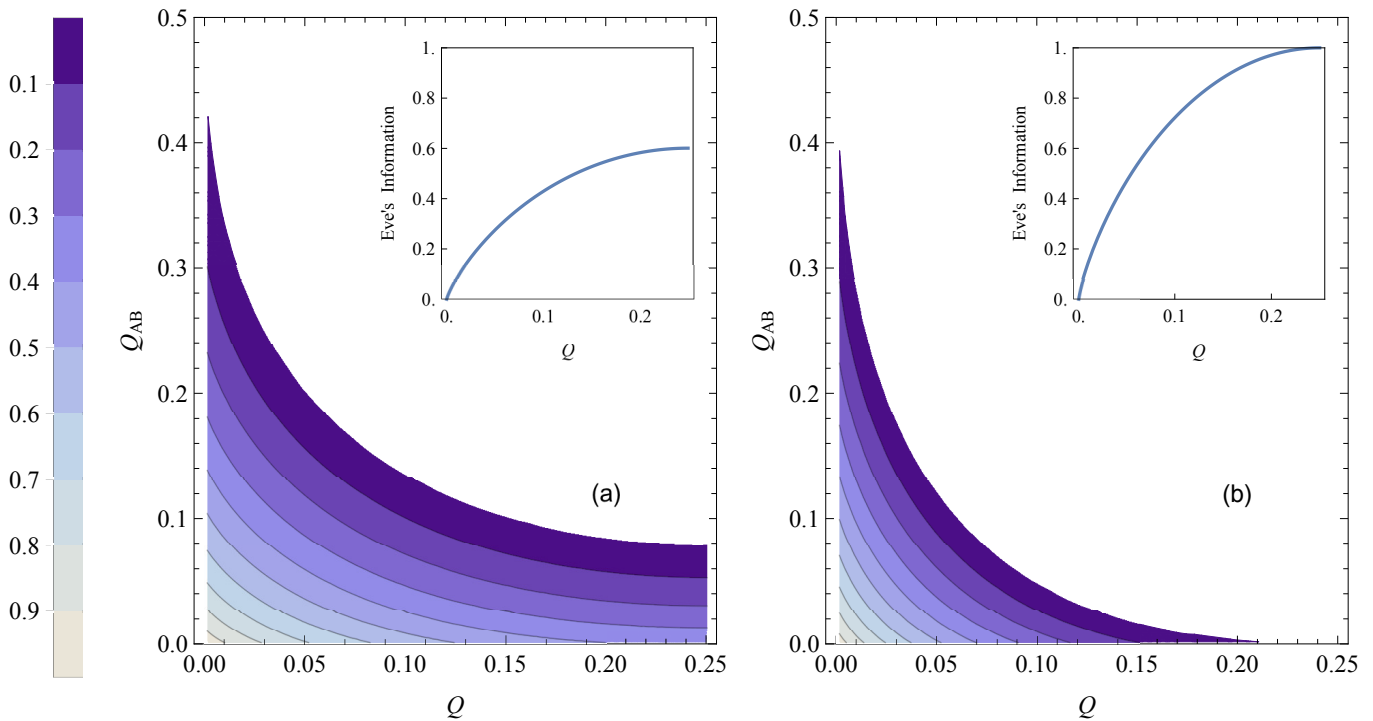} 
\caption{Comparison of secret key rate (per bit) as contour plots, denoting only the positive key rate regions and Eve's information gain (insets) for the current protocol (a) against LM05 (b). The insets show that the maximal amount of information Eve can gain (for maximal disturbance) in LM05 is $1$ (complete) while only about 0.6 for the current protocol.}
\label{test3}
\end{figure*}

The contour plot on the left, FIG. 1(a) represents the current protocol while FIG. 1(b) is for LM05. The insets shows Eve's gain for each protocol respectively. The figures clearly demonstrate how utilising these nonorthogonal transformations suppresses Eve's information, quite drastically in fact, as we observe the region (defined by $Q$ and $Q_{AB}$) for extractable secure key is greatest for the current protocol. This is mainly due to the lesser gain by an eavesdropper for the current protocol (insets). A direct numerical integration (using a mathematical software) gives the security thresholds for the current protocol and  LM05 
as $\approx 0.037$ 
and $ \approx 0.017$ respectively. It is perhaps instructive to note that had we compared absolute secure key rates, then a factor of $1/4$ would be multiplied to the key rate for the current protocol relative to that of LM05; this however does not change the security thresholds.

We now return to the issue of the inaccessibility of the value $Q$. As noted in \cite{bisio}, protocols like LM05 using orthogonal unitary transformation requires a CM. In principle, given the fact that Alice's encoding in the current protocol cannot be ascertained perfectly by Eve, even for a maximal attack ($Q=0.25$) the CM is, to a certain extent obsolete. A naive way of putting this would be to say that a key can still be distilled, without knowing $Q$ provided the error in Bob's (raw) key is less than a certain $Q_{AB}$ while assuming Eve has maximal information independent of errors in the raw key. We can simply calculate $Q_{AB}$ as follows: Alice and Bob can have a positive key rate provided $h(Q_{AB})< 1-\max{(I_E)}$, i.e. $Q_{AB}\approx 7\%$.  Thus in some sense, having a semblance of CM for the current protocol would only provide for a better key rate as Eve's information gain can be ascertained properly.

However, as we shall see shortly, practical considerations may delegate the estimation of $Q$ to a more critical role, especially given possible physical realisations with the use of polarisation of photons as qubits and waveplates for transformations.


\section{A Practical Protocol}

Let us consider a practical implementation of the protocol using the polarisation degree of photons as qubits. Realistic implementations of unitary transformations process pulses of photons independently of the actual number of photons. This of course exposes the protocol to a Quantum Man in the Middle attack where Eve could hijack Bob's photon en route and estimate Alice's transformations perfectly using a bright pulse before encoding Bob's photon accordingly to be submitted to him. The solution to this problem is the use of CM. The CM itself should of course involve a finite number of bases, say $n$, used for preparations and measurements; lest Alice and Bob would have a probability of approaching zero to agree a basis on in CM, $lim_{n\rightarrow \infty} 1/n = 0$.

We can thus imagine adding a step such that with probability $c$, Alice executes a CM where she would measure the incoming qubit in a basis selected from $n$ pre-agreed bases. Bob should then include these $n$ bases in his EM so that 
%
with probability $c/n$, a CM is successful and Alice and Bob may estimate errors in the forward path.\footnote{originally, some two-way protocols include an error check in the backward path as well. However, given that the security analysis does not provide for a parameter checking on the backward path, we do not consider such a check here.} This immediately provides Alice and Bob with a means to access $Q$ and thus estimate Eve's information gain and the security analysis of the previous section holds.\footnote{we do not however consider imperfections to the extent of having multiphoton pulses nor do we consider losses in channels}  For practical purposes, we set $n = 2$, corresponding to the conventional CM where the bases used are mutually unbiased. For simplicity, we further set Bob's number of basis in EM to be $2$ as well and correspond to the same bases for CM. 


\section{Experimental Proof of Concept}
In the following we report on an experimental implementation of the practical protocol described above where Bob uses polarisation of photons derived from only 2 MUBs, namely the $X$ (diagonal) basis and the $Z$ (rectilinear) basis for his preparations and measurements. The setup is basically a proof of principle with modest apparatus utilising polarised photons from attenuated laser pulses as qubits and half wave plates for the encoding process. These should be rather conventional; for example, the former is quite standard in QKD experiments or the latter for orthogonal/ nonorthogonal unitary implementation in \cite{Cere,non}.  While we do simulate the presence of Eve by introducing noise on the forward and backward paths (`artificial depolarisation' akin to that in \cite{Cere}), it is important to stress that this is not meant to be a full scale secure implementation. For example, rather than have Bob randomly select between bases for his qubits, we allow the protocol to be executed with Bob choosing one bases for a certain number of runs and another for the other runs; as is the case for Alice's encoding. We also do not execute classical aspect of a QKD protocol such as authenticating the users (Alice and Bob), error correction of Alice-Bob's strings and privacy amplification.

Figure 2 shows the schematic of the experimental setup which comprises of three main parts; namely Bob's, Eve's and Alice's sites.  Bob's site consists of a photon state preparation setup and Measurement Device (described in Methods) to analyse incoming photons in the backward path, whereas Alice's is composed of an Encoder (to be used in EM) and a Measurement Device (to be used in CM). 

\begin{figure*}
\centering \includegraphics[width=0.8\textwidth]{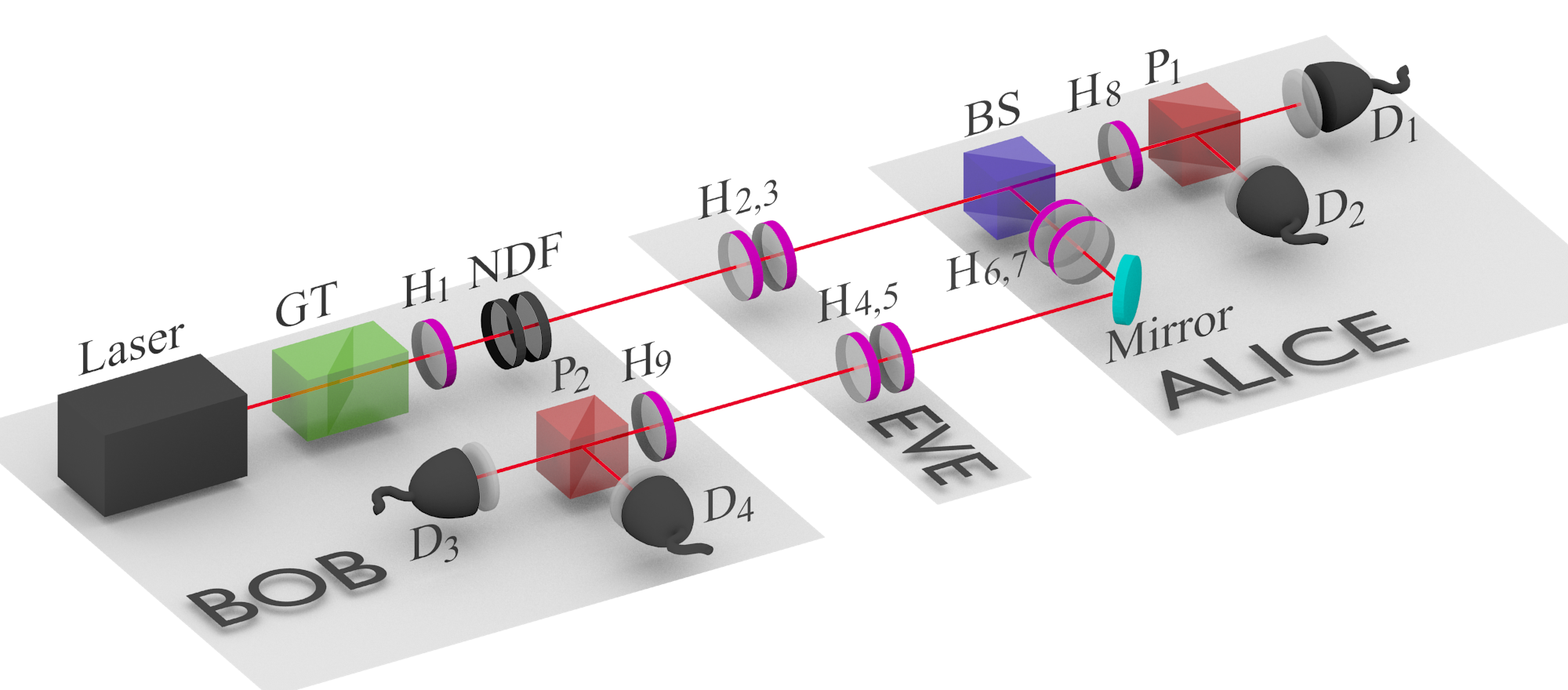} 
\caption{Sketch of the experimental setup. P$_i$ are Polarising Beamsplitters, D$_i$ are detectors (avalanched photodiode modules) and H$_i$ are Half waveplates (H$_{i,j}$ refers to H$_i$ and H$_j$). GT is a Glan-Thomson polariser, BS a beamsplitter and NDF are neutral density filters.}
\label{fig01}
\end{figure*}
At Bob's site, photon states were generated by a strongly attenuated laser and the polarisation of the photons can be set by using a zero-order half-wave plate. The polarised photon is then transmitted to Alice site via free-space (forward-path). 

Alice passively switches between CM and EM using a 50/50 beam splitter (BS). In CM, Alice would measure the incoming photon directly in the forward-path using a Measurement Device (described in the Methods section) with the waveplate (H$_8$). In the experiment, for the sake of simplicity, we always set Alice's measurement bases to be equal to Bob's preparation bases. Obviously a full fledged implementation would require Alice to measure in either the rectilinear or diagonal bases randomly where half would eventually be discarded.

In the EM mode, Alice would need to realise the $\mathbb{I}$ and $R_y(-\pi/2)$ operators. A pair of half wave plates (H$_{6,7}$) is used for the purpose before forwarding the qubit to a mirror, thus returning it to Bob via free space in  the backward path. 

The incoming polarisation encoded photons from Alice are finally analysed at Bob's site using a Measurement Device (refer to Methods) with the zero-order half waveplate (H$_9$) set in either the same bases the qubits were originally prepared or one rotated to the other basis.

The existence of the eavesdropper in the forward path and backward path is simulated by introducing noise in the communication system. This is done by virtue of `artificial depolarisation' channels in the forward as well as backward paths. It is important to note that we do not require the errors in the forward path to be equal to that in the backward nor do we require the errors in EM to be trivially related to that in CM; thus discarding the usual models of noise considered for two-way QKDs.

We use two independent personal computers equipped with PCIe field-programmable gate array card (FPGA; National Instruments PCIe-7853) to control and synchronise all active equipment in the experiment as well as for data acquisition purposes. 

\subsection{Experimental Results}

Data were collected for each `noise' setting; thus for each setting, we would have a pair of (average) errors, i.e. one from CM and the other from EM. In the case for EM, choosing only data which in principle would provide Bob with conclusive inference of Alice's encodings, we consider which of the cases tally with the actual encoding used by Alice and which do not (errors). Averaging the results over all states (by Bob) and encoding (by Alice) used, we arrive at an averaged error rate for the EM. 

Data providing error rates for CM is quite straightforward as we only compare the state sent by Bob to that measured in CM. As argued earlier, that one may not know \textit{a priori} the relationship between the errors in CM and EM (assuming there is one), we do not presume plotting the more conventional `information curves' for Eve's gain (which is a function of $Q$) and Alice-Bob's (which is a function of $Q_{AB}$). Instead, points are plotted based on these error rates as $Q_{AB}$ versus $Q$  as in FIG. 3. We include the contour lines of the previous FIG. 1 to exhibit which of the points fall below the security threshold and which beyond. The values accompanying the points are just the value for the corresponding (in theory) secret key rates calculated using the secret key rate formula. 

\begin{figure}[h]
\centering \includegraphics[width=\linewidth]{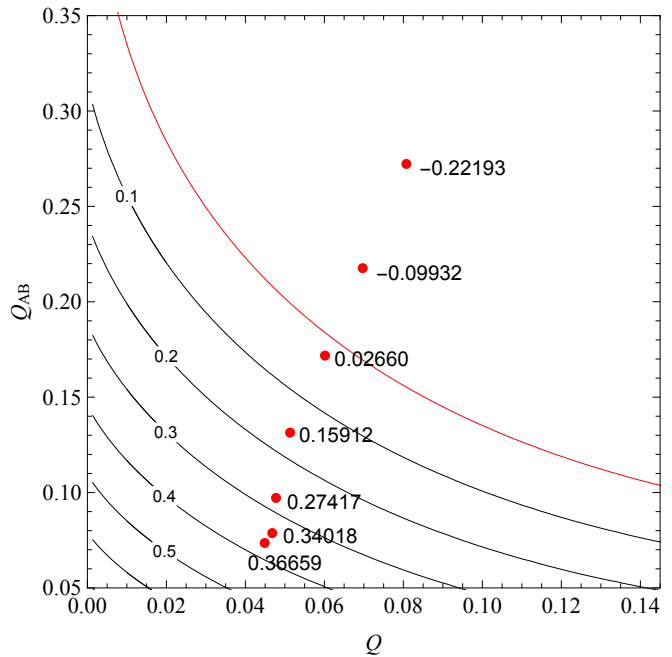} 
\caption{The red points are derived from the experiment. The numbers accompanying each point represent ``secure key rate" calculated using the key rate formula for the current protocol. The contour curves are based on earlier calculations as in FIG.1. The red curve represents the boundary between the region for distillable key (below) and non-distillable key (above).}
\label{fig02}
\end{figure}
While the points that occupy the secure region in the figure (positive secret key rate) reflect a very small sample of points that can, in principle be
achieved experimentally, we believe that these results already point out to the feasible and practical picture of our protocol. 


\section{Conclusions}

\noindent Bidirectional or two way QKD has certainly been a topic of interest for more than a decade now, ranging from entanglement based protocols, nonentangled versions and even continuous variables framework \cite{pirandola}. These protocols essentially have a common topology; where one party sends quantum states to another who would encode with a transformation before sending it back to the sender for his decoding measurement. Building on this, we demonstrate the simplest way forward for such protocols to actually embody the essence of creating an ambiguity for Eve to determine the encoding rather than rely strictly on the use of nonorthogonal states as information carriers to suppress her information gain. We thus proposed and analysed in this work, a novel bidirectional QKD protocol making use of only two nonorthogonal unitary transformations selected from two different MUUBs for encoding purposes along a rather different decoding procedure, akin to the SARG protocol \cite{SARG}.  Theoretical analysis based on collective attacks, coupled with a more relevant definition for security threshold has provided with a promising picture for the protocol's security compared to its predecessor. 

We have also executed an experimental setup for a proof of concept of the protocol using weak photon pulses traversing artificial depolarising channels between the legitimate parties with a pair of half wave plates for the encoding transformation. While we do not commit to actually distilling a secret key, we have demonstrated its feasibility given a very modest setup. A full scale protocol with actual secret key extraction is hence very possible given some addendum to the setup to include randomisation of Bob's choice for preparation and measurement, Alice's choice for encoding as well as a proper execution of error correction and privacy amplification protocols. 

Given its promising security, we hope that the future would see more realistic issues be addressed; a quick example would be the issue of multiphoton pulses coupled with channel losses and how the current protocol would perform given such imperfections. On a more fundamental note, we believe, this work should engender further interest, especially regarding the role of indistinguishable unitary transformations and even MUUBs within the context of quantum cryptography and quantum information as a whole.



\section{Methods}

In the following, we provide some details on the method we used in the theoretical calculations, mainly ascertaining the eigenvalues for the state $\rho_{BE}$ as well as certain detailed aspects of the experimental setup.

\subsection{Theoretical calculation for Eve's information}
We show in this section how one calculates the eigenvalues for the density operator $\rho_{BE}$. This is done, following \cite{ivan} by calculating the eigenvalues of its Gramm matrix representation. We start by noting the definition for the Gramm matrix representation \cite{gram} for an ensemble of pure states $\{ |\phi_i\rangle,...,|\phi_n\rangle,p_i,...,p_n\}$, a state given as $\sum_i^n|\phi_i\rangle\langle\phi_i|$ can be represented by a Gram matrix $\textbf{G}$ with the elements $G_{ij}=\sqrt{p_ip_j}\langle\phi_i|\phi_j\rangle$. Thus, to write out the Gramm matrix $\textbf{G}^{BE}$, for $\rho_{BE}$, we first write out all the possible (pure) states for Bob and Eve (after Alice's encoding) as
\begin{eqnarray}
|\Psi_0\rangle=|0\rangle|E_{00}\rangle+|1\rangle|E_{01}\rangle,\\\nonumber
|\Psi_1\rangle=|0\rangle|E_{10}\rangle+|1\rangle|E_{11}\rangle,\\\nonumber
|\Psi_2\rangle=R_y^E|\Psi_0\rangle,~|\Psi_3\rangle=R_y^E|\Psi_1\rangle,
\end{eqnarray}
and the Gramm matrix for the mixture $\sum_i|\Psi_i\rangle\langle\Psi_i|/4$, (each of the pure state $|\Psi_i\rangle$ is equiprobable) is given by 
\[
\begin{bmatrix}
   1       & 0 & 1/\sqrt{2} & \alpha/\sqrt{2} \\
     0       & 1 & -\alpha/\sqrt{2} & 1/\sqrt{2} \\
    1/\sqrt{2} & -\alpha/\sqrt{2} & 1 & 0\\
     \alpha/\sqrt{2} & 1/\sqrt{2} & 0 & 1
\end{bmatrix}
\]
with $\alpha=(F\cos{x}-Q\cos{y})$. The eigenvalues can then be easily calculated to be
\begin{eqnarray}
\lambda_\pm=\dfrac{1}{8}(2\pm \sqrt{2(F\cos{x}-Q\cos{y})^2+2}).
\end{eqnarray}
and hence
\begin{eqnarray}
S(\rho_{BE})=-2(\lambda_{+}\log_2{\lambda_{+}}+\lambda_{-}\log_2{\lambda_{-}}).
\end{eqnarray}
The above calculation can be repeated with Bob using the bases defined by the following states 
\begin{eqnarray}
|0^q\rangle&=&\cos{(\theta/2)}|0\rangle+\sin{(\theta/2)}|1\rangle,\\\nonumber
|1^q\rangle&=&\sin{(\theta/2)}|0\rangle-\cos{(\theta/2)}|1\rangle 
\end{eqnarray}
and we can write out the states of Bob and Eve subsequent to Alice's encoding as 
\begin{eqnarray}
|\Psi_0^q\rangle=|0^q\rangle|E_{00}^q\rangle+|1^q\rangle|E_{01}^q\rangle,\\\nonumber
|\Psi_1^q\rangle=|0^q\rangle|E_{10}^q\rangle+|1^q\rangle|E_{11}^q\rangle,\\\nonumber
|\Psi_2^q\rangle=R_y^E|\Psi_0^q\rangle,~|\Psi_3^q\rangle=R_y^E|\Psi_1^q\rangle
\end{eqnarray}
where Eve's ancillary states in the above equation are given by
\begin{eqnarray}
|E_{00}^q\rangle=\mathcal{C}_1|E_{00}\rangle+\mathcal{C}_2|E_{01}\rangle+\mathcal{C}_2|E_{10}\rangle+\mathcal{C}_3|E_{11}\rangle\\\nonumber
|E_{01}^q\rangle=\mathcal{C}_2|E_{00}\rangle-\mathcal{C}_1|E_{01}\rangle+\mathcal{C}_3|E_{10}\rangle-\mathcal{C}_2|E_{11}\rangle\\\nonumber
|E_{10}^q\rangle=\mathcal{C}_2|E_{00}\rangle+\mathcal{C}_3|E_{01}\rangle-\mathcal{C}_1|E_{10}\rangle-\mathcal{C}_2|E_{11}\rangle\\\nonumber
|E_{11}^q\rangle=\mathcal{C}_3|E_{00}\rangle-\mathcal{C}_2|E_{01}\rangle-\mathcal{C}_2|E_{10}\rangle+\mathcal{C}_1|E_{11}\rangle
\end{eqnarray}
with
$\mathcal{C}_1=\cos^2{(\theta/2)}$, $\mathcal{C}_2=\cos{(\theta/2)}\sin{(\theta/2)}$ and $\mathcal{C}_3=\sin^2{(\theta/2)}$.
This results in another Gramm matrix with equal eigenvalues. Thus Eve's information remains the same irrespective of Bob's choice of states. Further to that, setting
$\langle \mathcal{E}_{00}| \mathcal{E}_{00}\rangle=\langle \mathcal{E}_{00}^q| \mathcal{E}_{00}^q\rangle=F$ gives
\begin{eqnarray}
F=\dfrac{(1+\cos{y})}{2+(\cos{y}-\cos{x})}
\end{eqnarray}
This is just a rederiving of the same in \cite{Gisin}.

\subsection{Experiment}  

\noindent\textbf{Photon State Preparation}
Polarised photon states were generated by a strongly attenuated pulsed laser diode (Coherent, OBIS 785 LX) using variable neutral density filters (NDF). The use of variable NDF is to allow for the preparation of photon states with a certain averaged number per pulse, $\mu$. A Glan-Thomson polariser (GT), with an extinction ratio of 100000:1, was inserted after the attenuator to ensure linearly polarised photon states. A zero-order half-wave plate (H$_1$) after GT prepares particular polarised states; the polarisation states{$|z+\rangle$,  $|z-\rangle$}, and {$|x-\rangle$, $|x+\rangle$} were prepared by setting the polariser angles to $\varphi =0$, $\varphi =\pi/4$,  $\varphi =-\pi/8$, and  $\varphi =\pi/8$ with respect to  $z$-axis, respectively.  The density of photon $\mu$ was set to $\approx 0.15$ photon/pulse, measured just after the half-wave plate (H$_1$).
\\

\noindent\textbf{Measurement Device}
The measurement device is made up of  a set of zero-order half waveplate (H$_8$ for Alice and H$_9$ for Bob),  a polarizing beam splitter (P$_1$ for Alice and P$_2$ for Bob) and  two avalanched photodiode modules (APD) (D$_1$ and D$_2$ for Alice and D$_3$ and D$_4$ for Bob) with the quantum efficiency of about 70\% at a wavelength of 785nm. Incoming photons are collected into two multimode fibers using objective lenses (Newport M-10X, focal length 16.5mm, NA=0.25) through a pair of interference filters centered at 785nm with a bandwidth of 10nm (used to reduce background light) to eventually be detected by the APDs. 
\\

\noindent\textbf{Alice's Encoder}
Alice's encoding operation is realized by the use of a couple of half wave plates (H$_{6,7}$, for angles $\varphi_6,\varphi_7$ with respect to the $z$ axis) which rotates any state on the equator of the Poincare sphere by an angle $\varphi_6+\varphi_7$. The passive, $\mathbb{I}$ operation was realized by setting the angles of the half wave plates H$_{6,7}$ are to be $\varphi_6=\varphi_7  = 0$ radiant,  while the flipping of polarization states between the mutually orthonormal $Z$ and $X$ bases, $R_y(-\pi/2)$ is realised by setting the angles of H$_{6,7}$ to be $\varphi_6=\pi/8$ radian, and $ \varphi_7  = \pi/4$ radian.   
\\

\noindent\textbf{Artificial Depolarisation}
Artificial depolarisation is induced by simply inserting a pair of half-wave plates in the forward path (H$_{2,3}$) and another in the backward path (H$_{4,5}$). This has the effect (in each path) to rotate the state sent by Bob by a certain angle, thus resulting in a probability of an erroneous detection in the CM as well as EM. Varying levels of `depolarisation', or more precisely, erroneous detection in each path can be independently introduced by rotating the angles half-wave plates accordingly. It is worth noting that this artificial method while aimed at creating a symmetric error for the two bases used by Bob in CM, would create an asymmetry in the errors caused for different encodings used by Alice. However, we shall only be interested in the averaged error values for this experiment.

\section{Acknowledgement}

J. S. S. would like to thank Stefano Mancini (UNICAM) and Qing-yu Cai (WIPM) for helpful discussions and for financial support by the Ministry of Higher Education (Malaysia) under the Fundamental Research Grant Scheme FRGS14-152-0393. Both authors would like to thank Adhim (XMU) for his assistance in the figure related to the experimental setup (FIG. 2) and M. Ridza Wahiddin (IIUM) for his encouragement and support of the work.

\section{Author Contributions}
J. S. S. contributed to the theoretical part as well as the writing of the manuscript, while S. worked mainly on the experimental aspects as well as, to a lesser degree, the writing of the manuscript.

\end{document}